\documentclass{iaus}
\usepackage{graphicx}

\title[JD 11.~~The molecular gas in Luminous Infrared Galaxies] 
{The molecular gas in Luminous Infrared Galaxies: a new emergent picture}

\author[Papadopoulos,  Zhang,  Weiss et al.]   
{Padelis P. Papadopoulos$^1$, Zhi-Yu Zhang$^2$, Axel Weiss$^1$, Paul van der Werf$^3$, Kate Isaak$^4$,
Yu Gao$^2$, Manolis Xilouris$^5$, \& Thomas R. Greve$^6$}

\affiliation{$^1$ Max Planck Institute for Radioastronomy,  Auf dem H{\"u}gel 69, 
 D-53121 Bonn, Germany, email: {\tt padelis@mpifr-bonn.mpg.de}\\[\affilskip]
$^2$ Purple Mountain Observatory, Chinese Academy of Sciences, Nanjing, Jiangsu
 210008, China, email: {\tt pmogao@gmail.com}\\[\affilskip]
$^3$ Leiden Observatory,  Leiden University,  NL-2300 RA Leiden, 
The Netherlands, email: {\tt pvdwerf@strw.leidenuniv.nl}\\[\affilskip] 
$^4$ Research and Scientific Support Department, European Space Agency,
ESTEC, NL-2201, The Netherlands, email: {\tt kisaak@rssd.esa.int}\\[\affilskip]
$^5$ Institute for Astronomy, Astrophysics, Space Applications \&
Remote Sensing, National Observatory of Athens, P. Penteli,
15236 Athens, Greece, \\ email: {\tt xilouris@astro.noa.gr}\\[\affilskip]
$^6$ Department of Physics and Astronomy, University College London,
 London WC1E 6BT, UK,  email: {\tt tgreve@star.ucl.ac.uk}}

\pubyear{2013}
\volume{292}  
\pagerange{119--126}
\jname{Molecular Gas, Dust, and Star Formation in Galaxies }
\editors{Tony Wong, \& J{\"u}rgen Ott, eds.}
\begin{document}

\maketitle

\begin{abstract}
Results from  a large, multi-J CO,  $^{13}$CO, and HCN  line survey of
Luminous     Infrared    Galaxies     (LIRGs:     $\rm    L_{IR}$$\geq
$10$^{10}$\,L$_{\odot}$)  in the local  Universe ($\rm  z$$\leq $0.1),
complemented  by  CO  J=4--3  up  to J=13--12  observations  from  the
Herschel Space Observatory (HSO), paints a new picture for the average
conditions of the molecular gas of the most luminous of these galaxies
with  turbulence and/or large  cosmic ray  (CR) energy  densities $\rm
U_{CR}$ rather than far-UV/optical  photons from star-forming sites as
the   dominant   heating   sources.    Especially  in   ULIRGs   ($\rm
L_{IR}$$>$10$^{12}$\,L$_{\odot}$) the  Photon Dominated Regions (PDRs)
can encompass at  most $\sim $few\% of their  molecular gas mass while
the large  $\rm U_{CR}$$\sim $$10^{3}$\,$\rm U_{CR,  Galaxy}$, and the
strong turbulence in  these merger/starbursts, can volumetrically heat
much  of  their molecular  gas  to  $\rm T_{kin}$$\sim  $(100-200)\,K,
unhindered  by  the  high   dust  extinctions.   Moreover  the  strong
supersonic turbulence in ULIRGs  relocates much of their molecular gas
at  much higher  {\it  average} densities  ($\geq$10$^{4}$\,cm$^{-3}$)
than in isolated spirals ($\sim $(10$^2$--10$^{3}$)\,cm$^{-3}$).  This
renders low-J CO lines incapable of constraining the properties of the
bulk of the  molecular gas in ULIRGs, with  substantial and systematic
underestimates of  its mass  possible when only  such lines  are used.
Finally a  comparative study  of multi-J HCN  lines and CO  SLEDs from
J=1--0 up to J=13--12 of NGC\,6240 and Arp\,193 offers a clear example
of  two  merger/starbursts whose  similar  low-J  CO  SLEDs, and  $\rm
L_{IR}/L_{CO,1-0}$, $\rm  L_{HCN, 1-0}/L_{CO,1-0}$ ratios  (proxies of
the so-called  SF efficiency  and dense gas  mass fraction),  yield no
indications about their strongly diverging CO SLEDs beyond J=4--3, and
ultimately the  different physical conditions  in their molecular~ISM.
The  much larger sensitivity  of ALMA  and its  excellent site  in the
Atacama  desert now allows  the observations  necessary to  assess the
dominant energy  sources of  the molecular gas  and its mass  in LIRGs
without depending on the low-J CO~lines.


\end{abstract}

\section{Introduction}

The CO rotational lines are the most widely used probes of the average
conditions  and  mass  of  the  molecular  gas  in  galaxies,  with  a
substantial body  of data  assembled over the  last two  decades (e.g.
\cite[Braine   \&   Combes   1992]{BrComb92};   \cite[Aalto   et   al.
  1995]{Aa95};  \cite[Solomon et  al.   1997]{Sol97}, \cite[Downes  \&
  Solomon 1998]{DoSo98},  \cite[Mauersberger et al.   1999], \cite[Yao
  et  al.  2003]{Yao03},  \cite[Mao  et al.   2011]{Mao10}).  Most  of
these are  for J=1--0, 2--1, with  only few datasets  having J=3--2 or
higher-J lines.  The  lowest J=1--0 line serves as  a global molecular
gas  mass  tracer via  the  so-called  $\rm X_{CO}$=$\rm  M(H_2)/L^{'}
_{co,1-0}$ factor (\cite[Dickman  et al.  1986]{Dick86}; \cite[Solomon
  \&  Barrett 1991]{SolBar91}).   The low  excitation  requirements of
this  transition ($\rm E_1/k_B$$\sim  $5.5\,K and  $\rm n_{crit}$$\sim
$400\,cm$^{-3}$)  allow even  the coldest  and lowest  density  gas in
ordinary  Giant Molecular  Clouds (GMCs)  to have  a  substantial, and
hopefully calibratable, line luminosity contribution.

The next  two decades  were then spend  formulating the  dependance of
$\rm X_{CO}$ on the  average molecular gas conditions, and calibrating
its  values  for  various  ISM environments  (\cite[Maloney  \&  Black
  1988]{MalBla88},    \cite[Young    \&    Scoville    1991]{YoSco91},
\cite[Wolfire   et  al.    1993]{Wol93},   \cite[Bryant  \&   Scoville
  1996]{BrySco96}, \cite{DoSo98}).  Aside  from a strong dependance of
$\rm  X_{CO}$ on  metallicity  that could  leave  large reservoirs  of
molecular  gas in  the outer  parts  of spirals  or metal-poor  dwarfs
untraceable   by   CO   (\cite[Papadopoulos  et   al.    2002]{Pap02},
\cite[Wolfire et  al.  2010]{Wol10}),  any deviations of  $\rm X_{CO}$
from      its      Galactic      value     of      $\sim      $5\,$\rm
M_{\odot}$\,(K\,km\,s$^{-1}$\,pc$^2$)$^{-1}$   were  determined  using
radiative transfer  models of  CO lines to  constrain the  average gas
density   $\rm  n(H_2)$,   CO  J=1--0   brightness   temperature  $\rm
T_{b,1-0}$, and dynamical state upon which $\rm X_{co}$ depends~as

\begin{equation}
\rm X_{CO}=\frac{M(H_2)}{L^{'} _{co, 1-0}}=\frac{3.25}{\sqrt{\alpha}} 
\frac{\sqrt{n(H_2)}}{T_{b,1-0}} K_{vir}^{-1}\, \left(\frac{M_{\odot}}{K\,km\,s^{-1}\,pc^2}\right),
\end{equation}

\noindent
(\cite[Papadopoulos  et al.  2012a]{Pap12a})  where $\alpha$=0.55--2.4
depending on the cloud density profile, and $\rm L^{'} _{co,1-0}$=$\rm
\int     _{\Delta     V}\int_{A_s}     T_{b,1-0}da\,dV$     is     the
velocity/area-integrated  CO  J=1--0  brightness  temperature  at  the
reference  frame  of  the  source.   The parameter  $\rm  K_{vir}$  is
given~by

\begin{equation}
\rm K_{vir}=\frac{\left(dV/dR\right)}{\left(dV/dR\right)_{virial}}\sim 
1.54\frac{[CO/H_2]}{\sqrt{\alpha}\Lambda _{co}}\left(\frac{n(H_2)}{10^3\,
 cm^{-3}}\right)^{-1/2},
\end{equation}

\noindent
where  $\rm [CO/H_2]$$\sim  $10$^{-4}$ is  the CO  abundance  and $\rm
\Lambda _{co}$=$\rm  [CO/H_2]/(dV/dR)$ is one of  the three parameters
(the  other two  being $\rm  n(H_2)$ and  $\rm T_{kin}$)  of one-phase
Large Velocity  Gradient (LVG)  radiative transfer models  (with dV/dR
the gas velocity gradient).  The values of $\rm K_{vir}$ determine the
average gas  dynamical state, with $\rm K_{vir}$$\sim  $1-2 typical of
self-gravitating  (or  nearly  so)  states,  and  $\rm  K_{vir}$$\gg$1
indicating unbound  gas (\cite{Pap12a}).   In principle LVG  models of
the lowest  three CO lines can  produce constrains on  $\rm X_{CO}$ to
within  factors  of  $\sim  $2   via  Equation  1.1  (or  via  similar
expressions  in  the  literature,  e.g.   \cite{BrySco96})  since  the
excitation  characteristics  of   J=3--2  ($\rm  E_3/k_B$=33\,K,  $\rm
n_{crit}$$\sim  $10$^{4}$\,cm$^{-3}$) bracket the  conditions expected
for much  of the gas  in GMCs ($\rm T_{kin}$$\sim  $(10-30)\,K, n$\sim
$(500-10$^3$)\,cm$^{-3}$). For  merger/starbursts such studies yielded
a           low           $\rm           X_{co}$$\sim           $1$\rm
M_{\odot}$\,(K\,km\,s$^{-1}$\,pc$^2$)$^{-1}$
(\cite{Sol97,DoSo98,Yao03}),  which  was   then  widely  used  in  the
literature for similar galaxies at low or high redshifts.

The $\rm X_{co}$ dependance on  the physical conditions of GMCs brings
forth the question of what  determines them (and thus $\rm X_{co}$) in
LIRGs.    When  it   comes   to  the   all-important  thermal   state,
far-UV/optical  photons from  star-forming (SF)  sites have  long been
considered as the main heating source of the molecular gas and dust in
the Galaxy, with  PDRs containing $\sim $90\% of  its molecular gas in
Photon Dominated Regions (PDRs)  where photons determine chemistry and
thermal balance  (\cite[Hollenbach \& Tielens  1999]{HolTiel99}).  The
molecular ISM and  its corresponding line and dust  continuum in LIRGs
is then considered fully reducible to PDR ensembles (\cite[Wolfire
  et al.  1990]{Wolf90}).

\section{The  line survey of LIRGs: reaching out to the warm and dense gas}

We untertook  a large molecular line  survey of LIRGs,  drawn from the
\textit{IRAS}  Bright   Galaxy  Survey  (BGS)   (\cite[Soifer  et  al.
  1987]{Soi87};  \cite[Sanders et al.   2003]{San03}) using  the James
Clerk Maxwell  Telescope (JCMT) on Mauna  Kea in Hawaii,  and the IRAM
30-m  telescope on  Pico Veleta  in  Spain.  Our  line database,  when
augmented  by data  existing in  the literature,  yielded  the largest
multi-J dataset of {\it total} CO, $^{13}$CO and HCN line luminosities
for galaxies in the  local (z$\leq $0.1) Universe (\cite{Pap12b}). The
lines are  CO J=1--0, 2--1,  3--2, $^{13}$CO J=1--0 and/or  J=2--1, as
well as HCN  J=1--0, 3--2, and 4--3, while for  a smaller subsample CO
J=4--3  and J=6--5 are  also available.  This ground-based  effort has
been  complemented by  SPIRE/FTS  observations of  a  large number  of
ULIRGs in  our sample,  yielding complete CO  SLEDs from J=1--0  up to
J=13--12, and  an unparalled  view of the  densest and warmest  gas in
these merger/starbursts (see Figure 1).

\begin{figure}
\includegraphics[width=2.65in]{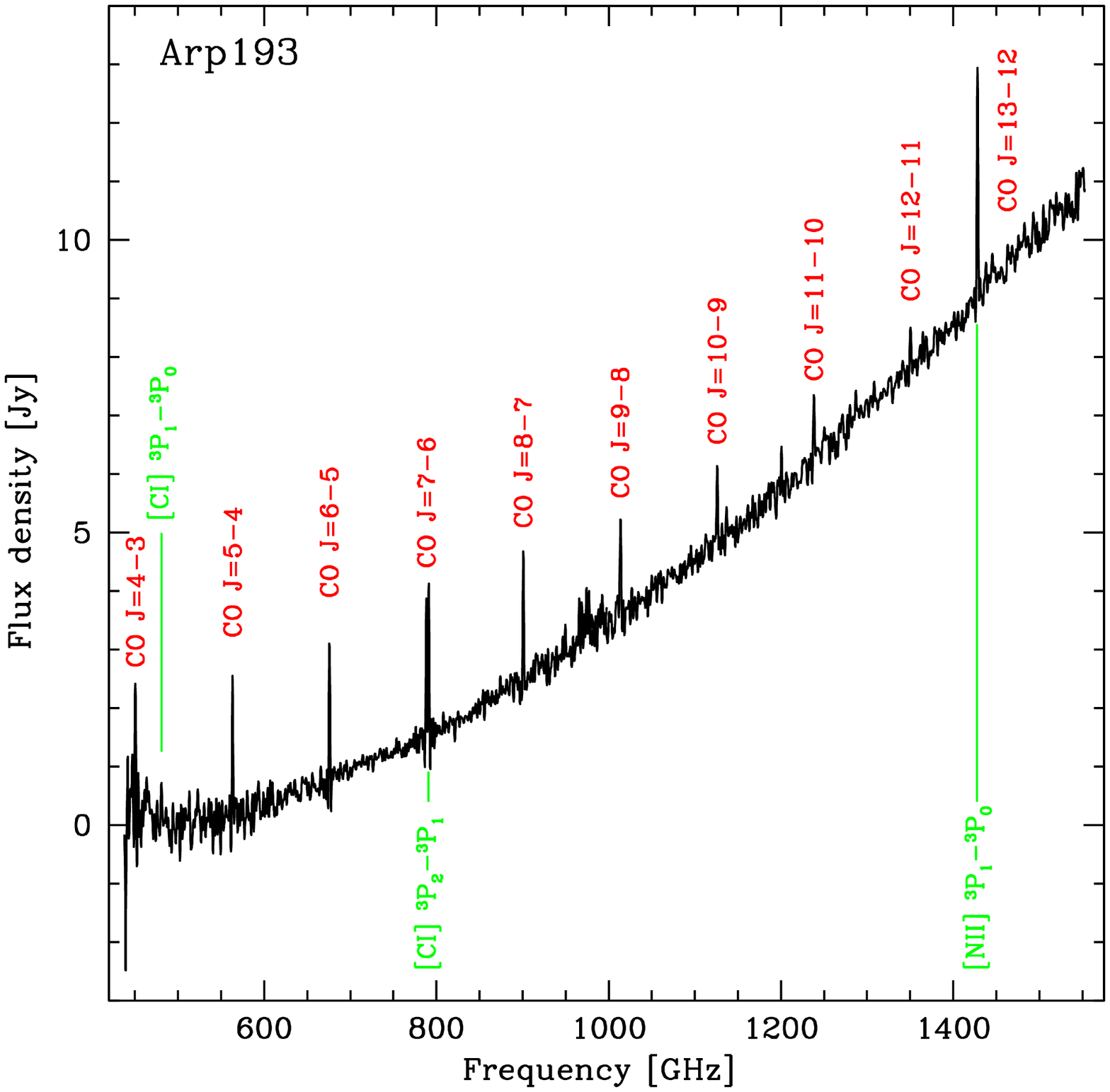}
\includegraphics[width=2.65in]{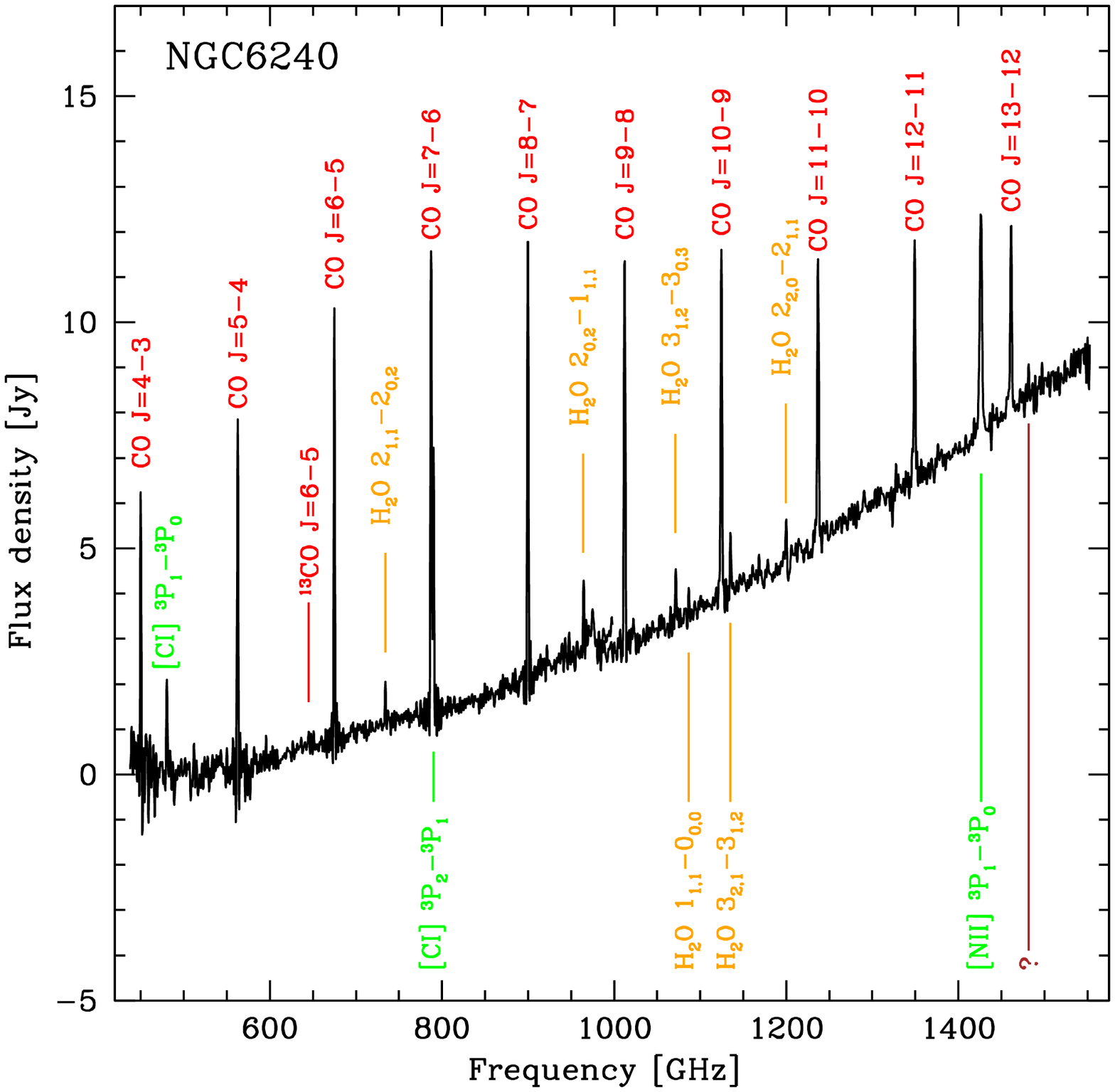}
\caption{The SPIRE/FTS spectra of Arp\,193, NGC\,6240. The 
  lines  are:  CO J=4--3  up to  J=13--12 (red),  the  two fine
  structure lines of [C\,I]$^{3}$$P _1$$\rightarrow $$^3$$P_0$ and $
  ^{3}$$P_2$$\rightarrow $$^3$$P_1$ and  [N\,II] (green).}
\label{fig1}
\end{figure}

Even  a  casual  inspection  of   the  high-J  CO  SLEDs  of  the  two
merger/starbursts  Arp\,193  and  NGC\,6240  (Figure  1)  reveals  two
systems  that,   despite  their  similar  low-J  CO   SLEDs  and  $\rm
L_{IR}/L_{CO,1-0}$, $\rm  L_{HCN, 1-0}/L_{CO,1-0}$ ratios  (proxies of
the so-called  SF efficiency and  dense gas mass fraction),  they have
strongly  divergent  CO   SLEDs  above  J=4--3,  indicating  different
properties  and/or mass  for their  warm  and dense  gas phase.   This
high-J CO SLED divergence appears  even more clearly in Figure 2 where
they are shown normalized by the corresponding IR luminosities.  It is
worth noting  that the CO/$^{13}$CO  J=2--1 line ratios for  these two
galaxies are also similar (\cite{Pap12b}).

Thus  {\it the  entire low-J  CO,  $^{13}$CO line  diagnostic that  is
  typically  used to  constrain the  average molecular  gas properties
  (and  the  corresponding  $\rm  X_{co}$  factor)  in  LIRGs  may  be
  inadequate for  the merger/starburst  systems among them.}   This is
not unexpected  since, unlike in  isolated SF disk galaxies,  the much
stronger supersonic  turbulence in the  molecular gas of  ULIRGs (e.g.
\cite{DoSo98}) will  relocate most of the resident  molecular gas mass
at average  densities of n$\geq $10$^{4}$\,cm$^{-3}$,  and thus beyond
the reach  of the low-J CO,  $^{13}$CO lines.  The latter  then can no
longer yield relevant corrections for the $\rm X_{co}$ factor in these
systems (\cite{Pap12a}).

\begin{figure}
\begin{center}
\includegraphics[width=3.6in]{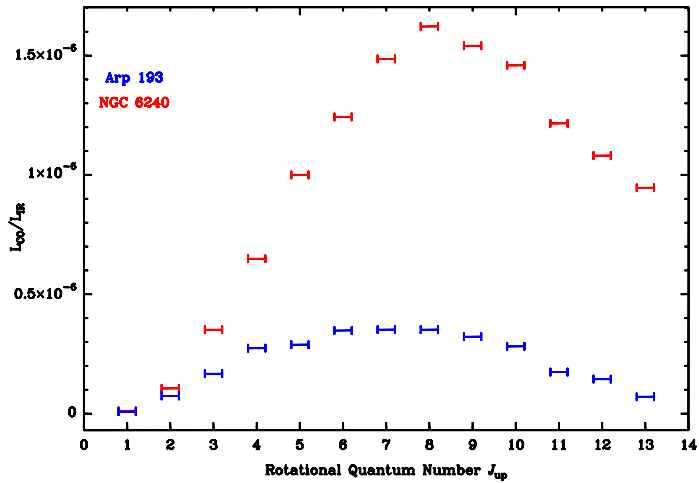}
\end{center}
\caption{The  CO SLEDs of Arp\,193 and NGC\,6240 normalized by their $\rm L_{IR}$
($\rm L_{\odot}/L_{\odot}$  units).}
\label{fig2}
\end{figure}

For Arp\,193 and NGC\,6240, the only ground-based observations able to
break   the  aforementioned   degeneracies  were   multi-J   HCN  line
observations  (\cite[Papadopoulos 2007]{Pap07}).   These found  a very
low HCN(4--3)/(1--0)  ratio in Arp\,193, compatible with  a total lack
of gas at  n$\geq $10$^{4}$\,cm$^{-3}$, quite unlike the  state of the
gas  in NGC\,6240,  and  what  is generaly  expected  in such  extreme
merger/starbursts   (\cite[Gao  \&   Solomon   2004]{GaSo04}).   These
diferences are seen  clearly in Figure 3 that  shows the $\rm [n(H_2),
  T_k]$  space  compatible with  the  HCN  line  ratios of  these  two
systems.  In the past  multi-J measurements of such high-dipole moment
molecules were impractical because of the limited sensitivities of the
available mm/submm telescopes.   This will no longer be  the case with
ALMA, and multi-J  HCN, HCO$^{+}$, CS observations of  (U)LIRGs may be
the  only way  of  assessing  the {\it  average}  conditions of  their
molecular gas  (and its mass) without  the aforementioned degeneracies
of  the  low-J CO  lines  (\cite{Pap12a}),  and  without depending  on
fully-sampled high-J  CO SLEDs. These will remain  difficult to obtain
past  J=6--5,  even  with  ALMA,  because  of  the  large  atmospheric
absorption at $\nu$$\geq$690\,GHz.

\begin{figure}
\begin{center}
\includegraphics[width=2.45in]{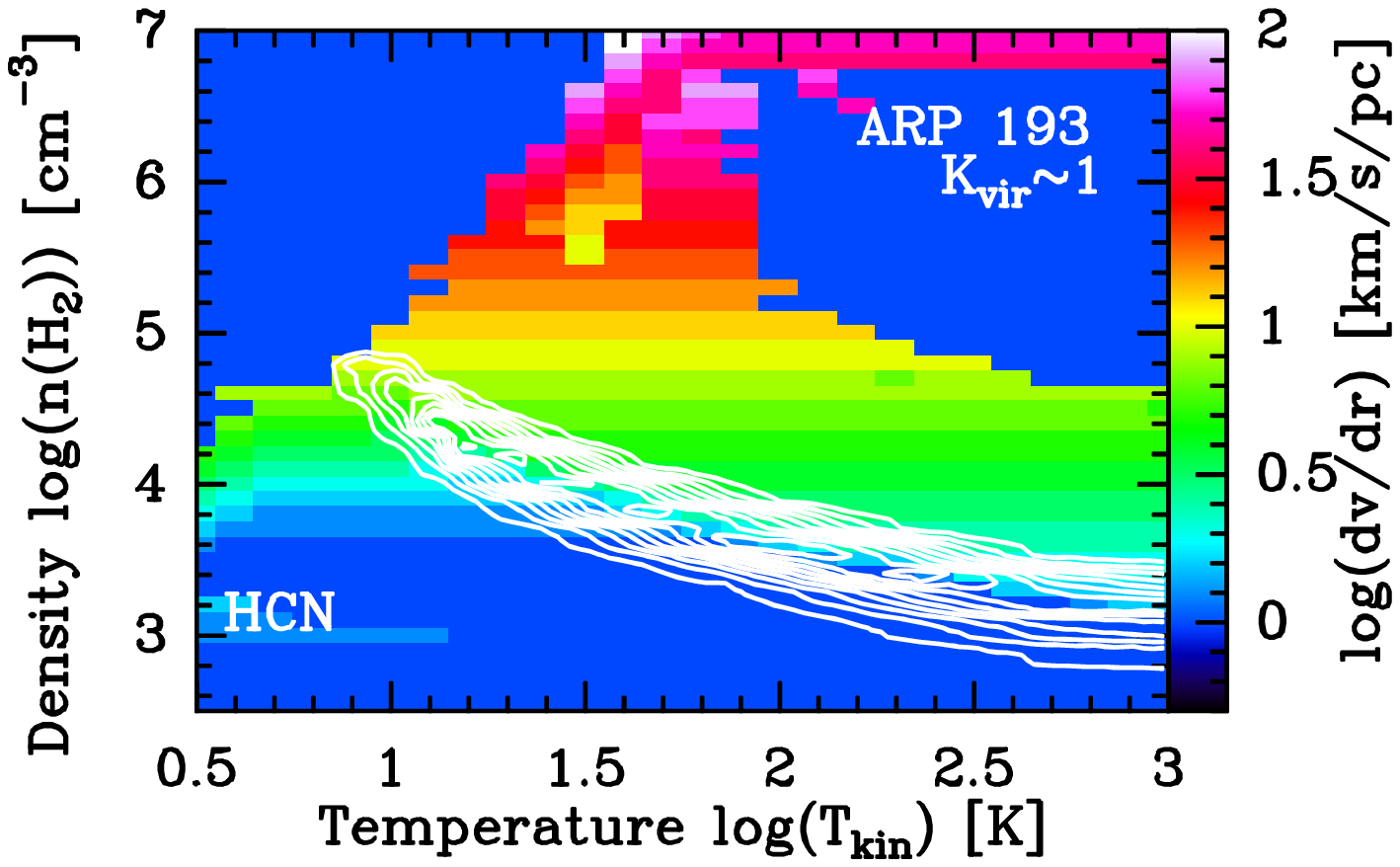}
\includegraphics[width=2.45in]{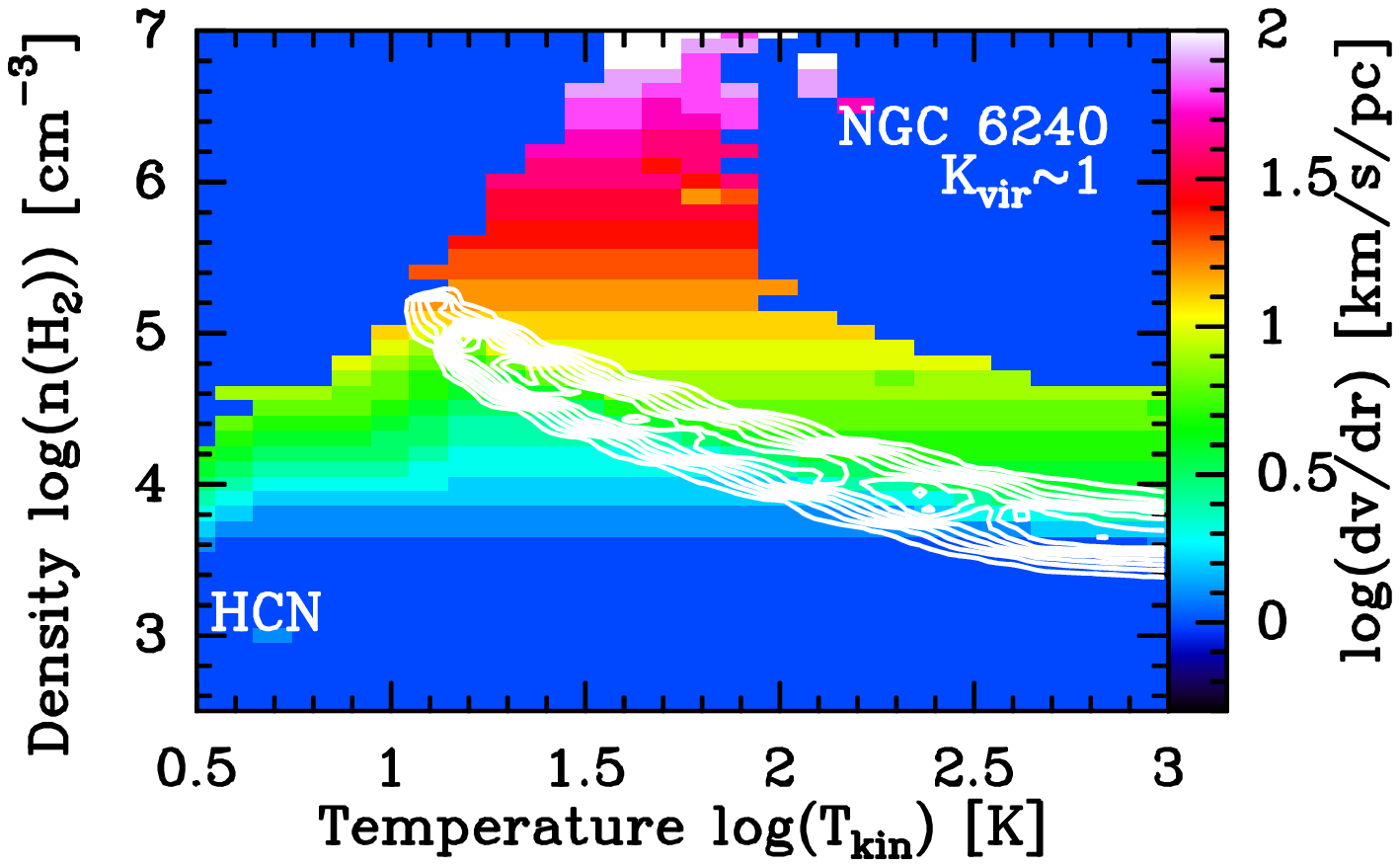}
\includegraphics[width=2.5in]{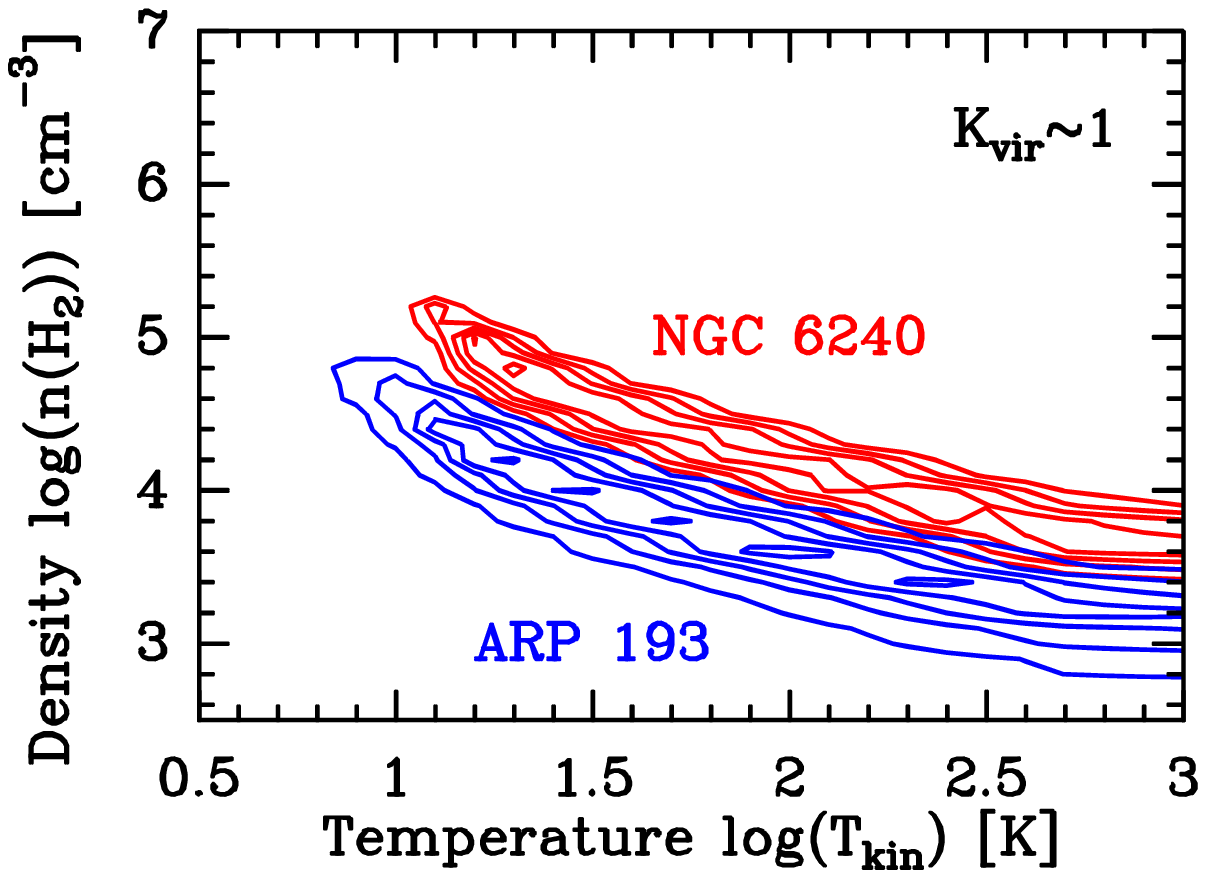}
\end{center}
\caption{The two-dimensional probability density functions of the $\rm
  [n, T_{kin}]$ LVG  solutions in steps of 0.2,  as constrained by the
  HCN line ratios of NGC\,6240 and Arp\,193.  Color: the corresponding
  $\rm (dV/dR)$ within the $\rm K_{vir}$=0.5--2 range (Equation 1.2).}
\label{fig3}
\end{figure}

\section{New power sources for the molecular gas in ULIRGs}

In  the high-density and  metal-rich molecular  ISM of  ULIRGs neither
far-UV/optical photons  nor SNR-induced shocks travel  far, the latter
dissipating strongly  in high density gas.  The  strong confinement of
warm PDRs and SNR-shocked regions  is evident even in the Galaxy where
most of  the molecular gas is  at much lower  average densities.  Such
regions involve  only $\sim  $(0.1-1)\% of the  mass of a  typical GMC
(\cite[Papadopoulos et  al.  2012b]{Pap12b}),  and this is  indeed why
even  the template  SF clouds  of Orion  A, B  are globally  cold with
CO(2--1)/(1--0)  ratios  of  $\sim  $0.6-0.7, and  only  few  isolated
regions near H\,II regions  reaching $\sim $0.9-1.3 (\cite[Sakamoto et
  al.   1994]{Sak94}).  The  corresponding global  CO(3--2)/(1--0) for
Orion would  then be  $\sim $0.30 while  for the  SF $''$hot-spots$''$
$\rm r_{32}$$\sim $0.9-1.3.  Thus it  is indeed a surprise that entire
galaxies can  approach and even surpass  these high-excitation regimes
(see   Figure   4).    These   excitation   outliers   are   typically
merger/starburst  LIRGs,  and  analysis  of  their  CO  ratios  (often
supplemented  by available CO  J=4--3, 6--5  lines) find  large ($\geq
$10\%) fractions of dense ($\sim $(10$^{4}$-10$^{5}$)\,cm$^{-3}$) {\it
  and} warm ($\sim $(100-200)\,K) molecular~gas.

\begin{figure}
\begin{center}
\includegraphics[width=2.2in]{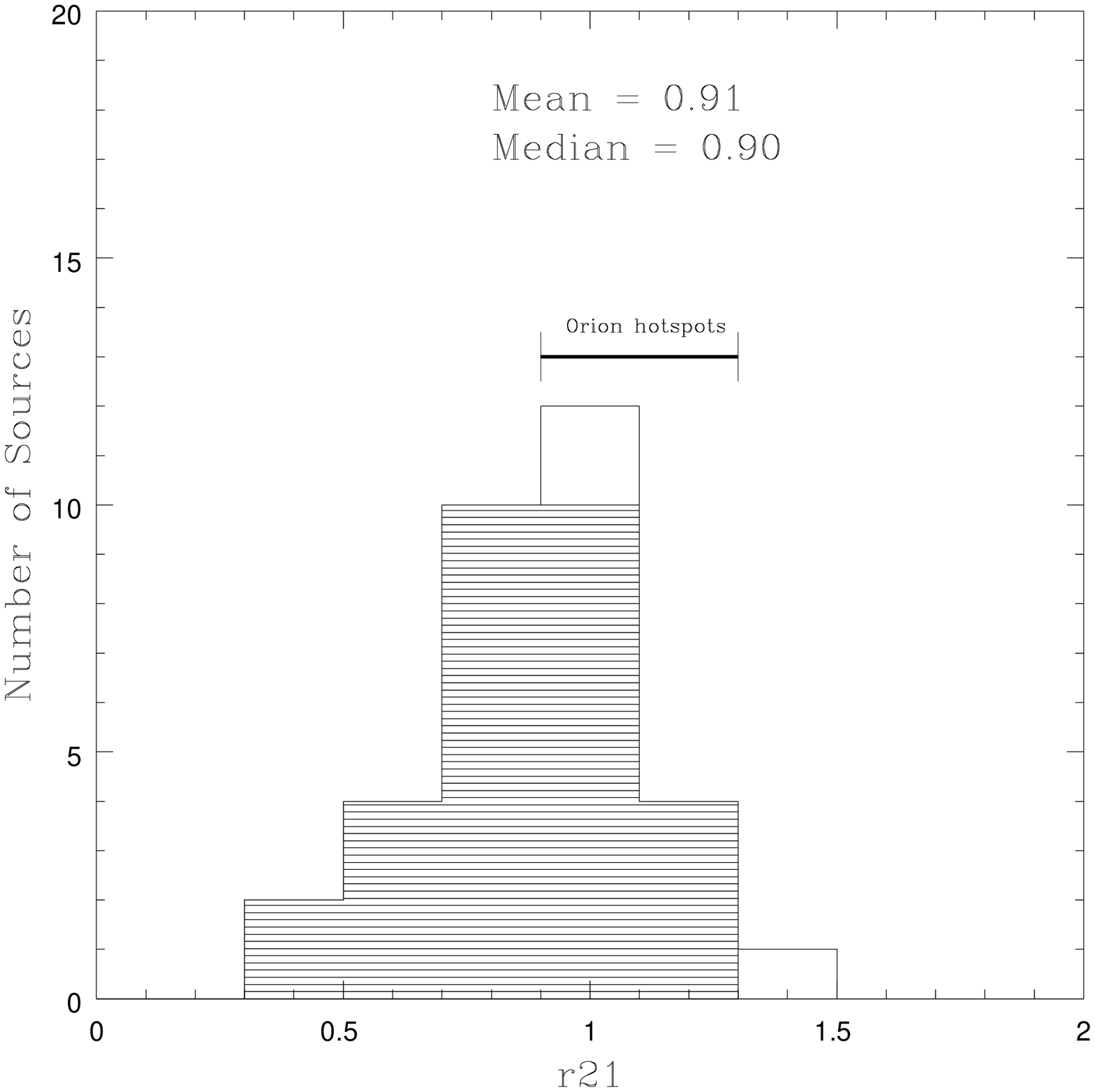}
\includegraphics[width=2.2in]{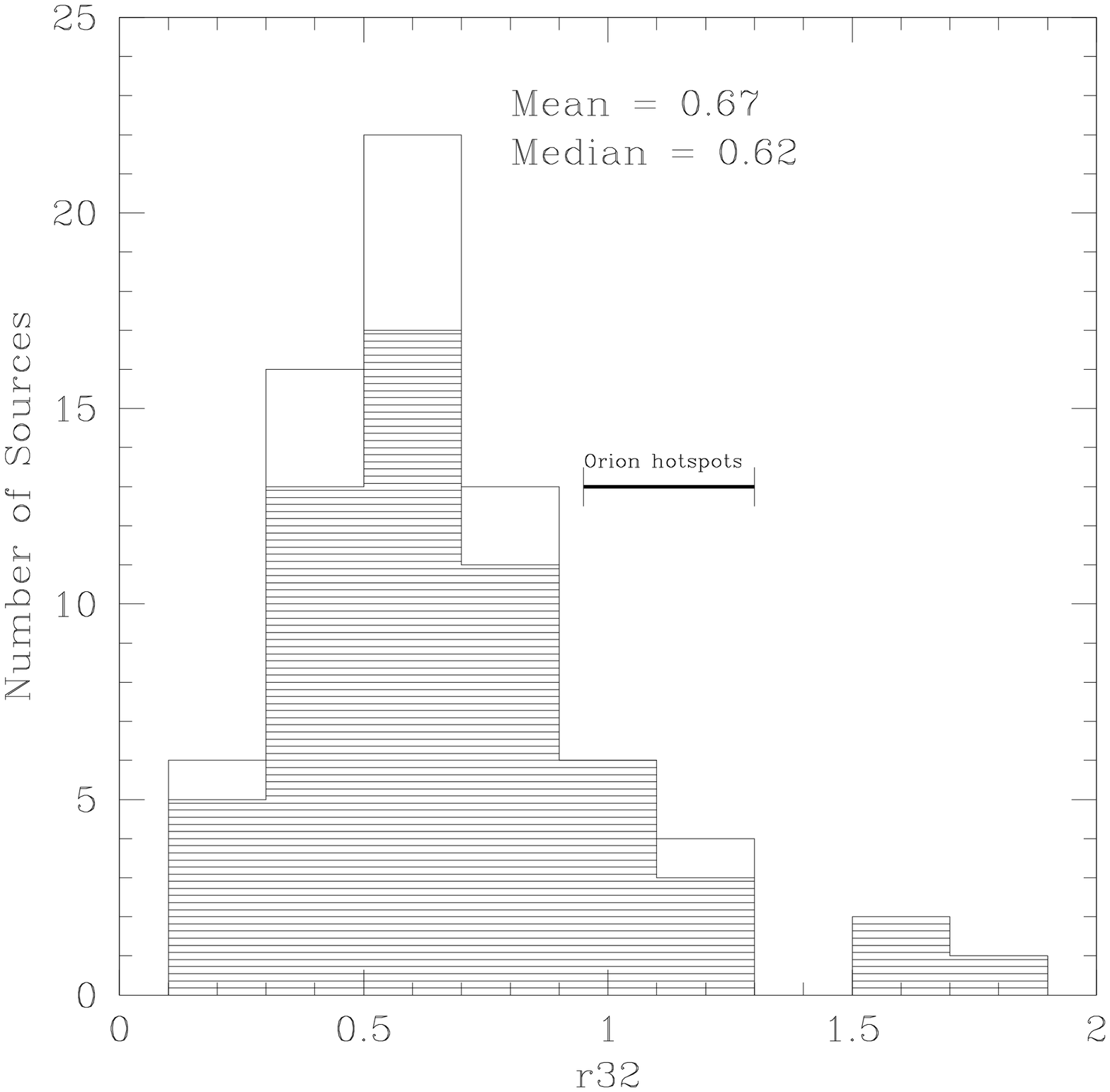}
\end{center}
\vspace*{-0.2cm}
\caption{The  distributions  of  the  CO  (2--1)/(1--0)  (3--2)/(1--0)
  brightness temperature  ratios for the galaxies in  the sample.  The
  shaded  area  marks  those   for  sources  with  CO  emission  sizes
  $\leq$15$''$  (the JCMT  beam).   The horizontal  bars indicate  the
  ratios for the Orion SF $''$hot-spots$''$.}
\label{fig4}
\end{figure}

For  NGC\,6240 and  Arp\,193 where  HCN multi-J  and fully  sampled CO
SLEDs  from  J=1--0  up  to   J=13--12  are  available  (Figure  1)  a
superposition of states from the HCN-constrained LVG solutions (Figure
3) with  $\rm T_{kin}$$\geq $70\,K  {\it is adequate to  reproduce the
  entire CO SLEDs up to  the highest transition.}  Only for the lowest
two  CO   lines  there  are  additional   contributions  from  unbound
low-density  gas   that  contains  little  mass.    For  NGC\,6240  in
particular  the  HCN-bright phases  that  can  reproduce its  luminous
high-J CO SLED with its  large line-continum contrast (Figs 1, 2) have
$\rm n(H_2)$$\sim $(few)$\times $10$^{4}$\,cm$^{-3}$, and contain most
of  its molecular  gas mass.   Far-UV photons  cannot drive  such high
temperatures  since  for  photoelectric  heating it  is:  $\rm  \Gamma
_{pe}$$\ll$$\rm   \Lambda_{CO}$+$\rm   \Lambda  _{H_2}$+$\rm   \Lambda
_{O\,I}$+$\rm  \Lambda _{gas-dust}$,  because the  line  cooling terms
(i.e.   CO, H$_2$  and O\,I  cooling) become  large at  high densities
($\rm \Lambda  _{line}$$\propto$$\rm [n(H_2)]^2$).  On  other hand the
strong supersonic turbulence  and the high CR energy  densities in the
ISM environment  of (U)LIRGs like NGC\,6240 can  easily maintain large
amounts of warm {\it and} dense molecular gas mass (\cite{Pap12a}).

\section{A Galactic X$_{\bf CO}$ in ULIRGs?}

Using Equation  1.1 with inputs from the  HCN-constrained LVG solution
space of NGC\,6240 shown in Figure 3 (which also satisfies the multi-J
CS and HCO$^{+}$  available for this LIRG) we  obtain the $\rm X_{CO}$
factor for the prevailing conditions  of its molecular gas (Figure 5).
It  is  obvious  that  for  the high-density  molecular  gas  of  this
merger/starburst  {\it  Galactic   $X_{CO}$ values  are  possible.}
Using the  SPIRE/FTS data  to select the  sub-regions of  the solution
space that can reproduce also the high-J CO lines selects regions with
$\rm  T_{kin}$$\geq $70\,K  that still  include Galactic  $\rm X_{CO}$
values, and point towards what may be a systematic understimate of 
molecular gas mass in ULIRGs (\cite{Pap12a}).

\begin{figure}
\begin{center}
\includegraphics[width=3.4in]{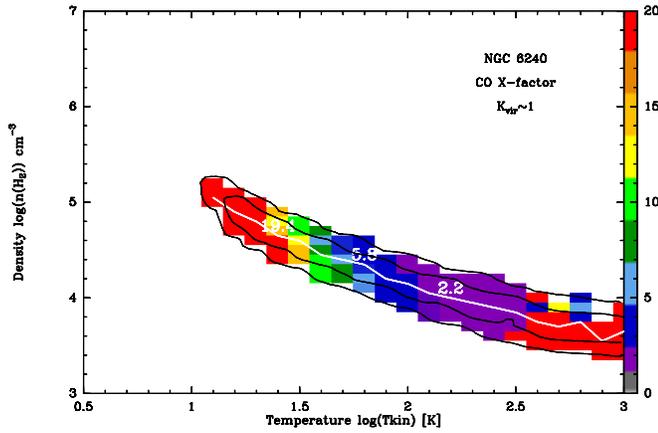}
\end{center}
\vspace*{-0.2cm}
\caption{The $\rm X_{CO}$ factor for NGC\,6240 and the conditions shown in Figure 3.}
\label{fig4}
\end{figure}

\end{document}